\documentclass[prd,aps,twocolumn,nofootinbib]{revtex4}
\usepackage[latin1]{inputenc}
\usepackage[english]{babel}
\usepackage{graphicx}
\usepackage{color}
\usepackage{amsmath}
\usepackage{amssymb}
\usepackage{epstopdf}
\usepackage{multirow}
\begin{document}
%
%
%
\title{\bf Black holes with constant topological Euler density}
\author{Pedro Bargue\~no$ ^{1}$} \email[]{p.bargueno@uniandes.edu.co}
\author{Elias C. Vagenas$^{2}$}\email[]{elias.vagenas@ku.edu.kw}
\affiliation{$ ^{1}$~Departamento de F\'{\i}sica,
Universidad de los Andes, Apartado A\'ereo {\it 4976}, Bogot\'a, Distrito Capital, Colombia}
\affiliation{$^2$~Theoretical Physics Group, Department of Physics, Kuwait University, P.O. Box 5969, Safat 13060, Kuwait}

\begin{abstract}
\par\noindent
 A class of four dimensional spherically symmetric and static geometries with constant topological Euler density is studied.
These geometries are shown to solve the coupled Einstein-Maxwell system when non-linear Born-Infeld-like 
electrodynamics is employed.
\end{abstract}

\maketitle
%
%
\section{Introduction}
%
%
%
%
%
%
\par\noindent
Since the discovery of Bekenstein  \cite{Bekenstein1972,Bekenstein1973,Bekenstein1974} and Hawking \cite{Hawking1974,Hawking1975}
connecting the entropy ($S$) and area ($\mathcal{A}$) of a black hole (BH), general relativity, quantum field theory and statistical
physics became deeply linked. Not only in BHs does  this connection work but also, as shown by Gibbons and Hawking 
\cite{Gibbons1977a}, the cosmological
horizon which is present in the de Sitter space fulfills $S=\frac{\mathcal{A}}{4}$. The geometric features of BH entropy seem to imply that it is related to the
non-trivial topological structure of the corresponding spacetime. Moreover, Hawking and Gibbons \cite{Gibbons1977b}
argued that, due to the different topologies between extremal and non-extremal BHs, the area law fails for the former because their entropy
vanish despite the non-zero area of the horizon. Furthermore, Teitelboim confirmed \cite{Teitelboim1995}, using the Hamiltonian formalism,
that extremal BHs had zero entropy. These facts led Liberati and Pollifrone \cite{Liberati1997a}
to suggest the formula $S=\frac{\chi \mathcal{A}}{8}$, where $\chi$ is the Euler characteristic of the corresponding regular and Riemannian version of the
considered spacetime (gravitational instanton). Although this formula has been checked for a wide class of gravitational instantons
\cite{Liberati1997a,Wang2000,Ma2003}, it is well know that it does not work in case of multiple horizon spacetimes \cite{Cai1998}.
\par\noindent
The Euler characteristic of a gravitational instanton is related with a particular combination of curvature invariants of the form
$\mathcal{G}=R^{\mu\nu\rho\sigma}R_{\mu\nu\rho\sigma}-4 R^{\mu\nu}R_{\mu\nu}+R^{2}$ by means of the Gauss-Bonnet (GB) theorem. This GB
term, $\mathcal{G}$, not only plays an important role in BH entropy but also in higher dimensional extensions of general relativity 
as well as in certain extensions in the four dimensional case. Specifically, the so-called $\mathcal{F}(\mathcal{G})$-gravity \cite{Nojiri2005}, where $\mathcal{F}$ is a
non-linear function of the four dimensional GB invariant, produces late-time accelaration \cite{Nojiri2006} and is an interesting alternative to standard
cosmology (see, for example, Ref. \cite{Felice2009} and references therein). Moreover, topological spherically symmetric vacuum solutions in $\mathcal{F}(R,\mathcal{G})$-gravity have been
recently studied \cite{Myrzakulov2013}.  Particularly, in Ref. \cite{Myrzakulov2013}, the authors looked for non-linear deformations of four dimensional
$\mathcal{G}$- and $R$-gravity theories such that solutions with a constant $\mathcal{G}$ appear.
In addition, the GB term is related to the trace anomalies in gravity (see Ref. \cite{Duff1994} for a review). 
Therefore, as pointed out in Ref.  \cite{Gibbons1995},
the knowledge of the GB term may be also useful to know how this quantum anomaly affects the classical solutions.
\par
In this work, a different intepretation of the static and spherically symmetric geometries supporting a constant $\mathcal{G}$ will be given
in terms of non-vacuum solutions of standard Einsteinian gravity. Our interpretation is somewhat similar in spirit
to that of Ref. \cite{Myrzakulov2013} but coupling certain model of non-linear electrodynamics (NLED) or null dust fluids (which generalize the
Vaidya solutions \cite{Vaidya1951}) to gravity, instead of deforming the gravitational action.
%
%
%
\section{Preliminaries: matter content}
%
%
%
\subsection{Non-linear electrodynamics}
%
%
%
\par\noindent
In geometrized units, Einstein equations ($\Lambda =0$) read
\begin{equation}
\label{einstein}
R_{\mu\nu}-\frac{1}{2}R\, g_{\mu\nu} = 8 \pi T_{\mu\nu},
\end{equation}
where $T_{\mu\nu}$ is the energy-momentum tensor.
\par\noindent
For the matter content we choose non-linear electromagnetic fields. To justify the study of these NLED theories, let us focus on two arguments.
First of all, quantum corrections to Maxwell theory can be described by means of non-linear effective Lagrangians that define NLED as, for instance,
the Euler-Heisenberg Lagrangian \cite{HE,Sch}. When higher order corrections are taken into account, we are led to a sequence of effective Lagrangians which
are polynomials in the field invariants \cite{Bia}. Among all the non-linear generalizations of Maxwell electrodynamics, Born-Infeld (BI) theory \cite{BI} has
been widely studied.  Interestingly, the BI Lagrangian depends on the two field invariants in the same way as
the one-loop effective Lagrangian for vacuum polarization due to a constant external electromagnetic field, which gives support to it.
A second argument comes from the low-energy limit of string theory. Specifically, in case of dealing with open bosonic strings, the resulting tree-level
effective Lagrangian is shown to coincide with the BI Lagrangian \cite{PLB1985,NPB1997}.
\par\noindent
When coupled to gravity, NLED gives place to interesting phenomena. The corresponding solutions give place to generalizations of the
Reissner-Nordstr\"om geometry, which have received considerable attention recently. In particular, BI solutions were presented
in \cite{Garcia1984,Breton2003}. BH solutions for generalized BI theories were
studied in \cite{Hendi2013}. An exact regular BH geometry in the presence of NLED was obtained in \cite{Ayon1998} and further discussed in
\cite{Baldo2000,Bronni2001}. Finally, BHs with the Euler-Heisenberg effective Lagrangian as a source term were examined in \cite{Yajima2001},
and the same type of solutions with Lagrangian densities that are powers of Maxwell's Lagrangian were analyzed in \cite{Hassaine2008}.
\par\noindent
In this work we consider a simple choice for an energy-momentum tensor for NLED which is written as
\begin{equation}
\label{nlT}
T^{\mu\nu}=-\frac{1}{4\pi}\left[\mathcal{L}(F)g^{\mu\nu}+\mathcal{L}_{F}F^{\mu}_{\;\;\rho}F^{\rho \nu} \right]
\end{equation}
where $\mathcal{L}$ is the corresponding Lagrangian, $F=-\frac{1}{4}F_{\mu\nu}F^{\mu\nu}$, and $\mathcal{L}_{F}=\frac{d\mathcal{L}}{dF}$ (along this work,
dependence on the second field invariant, $\sqrt{-g}/2\epsilon_{\mu\nu\rho\sigma}F^{\rho\sigma}F^{\mu\nu}$, will not be considered).
\par\noindent
For simplicity, let us take spherically symmetric and static solutions to Eqs. (\ref{einstein}) given by
\begin{equation}
\label{metric}
ds^{2}=-f(r)dt^{2}+f(r)^{-1}dr^{2}+r^{2} d\Omega^{2}.
\end{equation}
\par\noindent
In the electrovacuum case, we consider only a radial electric field as the source, namely,
\begin{equation}
\label{eqmaxwell}
F_{\mu \nu}=E(r)\left(\delta^{r}_{\mu}\delta^{t}_{\nu}-\delta^{r}_{\nu}\delta^{t}_{\mu} \right)
\end{equation}
where  $t$ and $r$ stand for  the time and radial coordinates, respectively. Maxwell equations will now read
\begin{equation}
\nabla_{\mu}\left(F^{\mu\nu}\mathcal{L}_{F}\right)=0,
\end{equation}
thus,
\begin{equation}
\label{max}
E(r)=-\frac{q}{r^{2}}(\mathcal{L}_{F})^{-1}.
\end{equation}
As pointed out in Ref. \cite{Cherubini2002}, the non-Weyl part of the curvature determined by the matter content can be separated by showing that
\begin{equation}
\label{main}
4 R^{\mu\nu}R_{\mu\nu}-R^{2}=(16 \pi)^{2}\left(T^{\mu\nu}T_{\mu\nu}-\frac{T^2}{4} \right)
\end{equation}
where $T=g^{\mu\nu}T_{\mu\nu}$ is the trace of the energy-momentum tensor. Therefore, in the considered spherically 
symmetric and static case, we arrive to the following expression for the electric field
\begin{equation}
\label{electric}
E(r)=\frac{r^{2}}{4q}\sqrt{4 R^{\mu\nu}R_{\mu\nu} -R^{2}}.
\end{equation}
It is important to point out that Eq. (\ref{electric}) is also valid for $\Lambda\ne0$, as can be easily shown by direct calculation.
\par\noindent
Let us now check Eq. (\ref{electric}) in two particular cases.
In the first case, we take a Reissner-Nordstr\"om BH whose relevant curvature invariants are given by
$R^{\mu\nu}R_{\mu\nu}=\nobreak 4 q^{4}/r^{8}$ and $R=0$. 
Therefore, $4 R^{\mu\nu}R_{\mu\nu}-R^{2}=16 q^{4}/r^{8}$ and $E(r)=q/r^{2}$, as expected.
\par\noindent
In the second case, let us consider a regular BH metric  \cite{Balart2014} given by $f(r)=1-\frac{2M}{r}e^{-\frac{q^{2}}{2Mr}}$.
For this geometry, we get  $R=e^{-\frac{q^2}{2 M r}} q^4 /2 M r^5$ and
$R^{\mu\nu}R_{\mu\nu}= e^{-\frac{q^2}{M r}} q^4 \left(q^4-8 M q^2 r+32 M^2 r^2\right)/ 8 M^2 r^{10}$.
Therefore, $4 R^{\mu\nu}R_{\mu\nu}-R^{2}=e^{-\frac{q^2}{M r}} q^4 \left(q^2-8 M r\right)^2 /4 M^2 r^{10}$ and
$E(r)=\frac{q}{r^{2}}\left(1-\frac{q^{2}}{8 M r} \right)e^{-\frac{q^{2}}{2 M r}}$ which coincide with Eq. (19) 
of  Ref. \cite{Balart2014}.
\par\noindent
The underlying NLED theory can be obtained using the $P$ framework \cite{Salazar1987}, which is somehow dual to the $F$ framework. After introducing the
tensor $P_{\mu\nu}=\mathcal{L}_{F}F_{\mu\nu}$ together with its invariant $P=-\frac{1}{4}P_{\mu\nu}P^{\mu\nu}$,
one considers the Hamiltonian-like quantity
\begin{equation}
\label{H}
\mathcal{H}=2 F \mathcal{L}_{F} -\mathcal{L}
\end{equation}
as a function of $P$, which specifies the theory. Therefore, the Lagrangian can be written as a function of $P$
as
\begin{equation}
\label{L}
\mathcal{L}=2 P \frac{d\mathcal{H}}{d{P}}-\mathcal{H}. 
\end{equation}
Finally, by reformulating the coupled Einstein-NLED equations
in terms of $P$, $\mathcal{H}(P)$ is shown to be given by \cite{Bronnikov2001}
\begin{equation}
\mathcal{H}(P)=-\frac{1}{r^2}\frac{d \mathcal{M}(r)}{dr}
\end{equation}
where the mass function $\mathcal{M}(r)$
is such that $f(r)=1-\frac{2 \mathcal{M}(r)}{r}$.
%
%
%
\subsection{Null fluids}
%
%
%
\par\noindent
Let us consider the geometry in terms of (ingoing) Eddington-Finkelstein coordinates
\begin{equation}
\label{genV}
ds^{2}=-f(r, v)dv^{2}+2dr dv+r^{2}d\Omega^{2}
\end{equation}
%
with $v=t+r$.
The kind of metrics represented by Eq. (\ref{genV}) are
a generalization of the Vaidya metric \cite{Vaidya1951} 
describing a spherically symmetric and nonrotating body which is either emitting ($u$-coordinate) or 
absorbing ($v$-coordinate) null dusts, i.e., ``incoherent" electromagnetic radiation. 
Moreover, it can be seen \cite{Husain1996,Wang1999} that they solve the Einstein equations
provided the associated energy-momentum tensor is that of a Type II perfect fluid \cite{book} given by
\begin{equation}
T_{\mu\nu}=(\rho + p)(l_{\mu}n_{\nu}+l_{\nu}n_{\mu})+p g_{\mu\nu}
\end{equation}
where $l_{\mu}=\delta^{t}_{\mu}$ and $n_{\mu}=\frac{1}{2}f(r)\delta^{t}_{\mu}-\delta^{r}_{\mu}$ are two 
null vectors. The density and pressure of the fluid are given by
\begin{eqnarray}
\rho&=&\frac{\mathcal{M}_{r}}{4\pi r^2} \nonumber \\
p&=&-\frac{\mathcal{M}_{rr}}{8\pi r}
\end{eqnarray}
where $\mathcal{M}_{r}\equiv d \mathcal{M}(r)/dr$.

%
%
%
\par\noindent
At this point, it is noteworthy that for Type II fluids, the energy conditions
read: (a) weak: $\rho\ge 0$ and $p+\rho \ge0$, 
(b) strong: $p\ge 0$ and $p+\rho \ge0$, and 
(c) dominant: $\rho\ge 0$ and $-\rho\le p \le \rho $.
%
%
%
%
%
%
%
%
\section{Solutions with constant topological Euler density}
%
%
%
%
\par\noindent
As mentioned in Refs. \cite{Nojiri2005,Nojiri2006, Myrzakulov2013},  several interesting cosmological models and 
topological static spherically symmetric solutions in $\mathcal{F}(R,\mathcal{G})$ gravity  are obtained 
when the topological Euler density, i.e., $\mathcal{G}$, is constant. In addition, in  $\mathcal{F}(\mathcal{G})$ gravity, 
during the transition from the matter to the accelerated era the topological Euler density is zero \cite{Felice2009}. 
For these reasons as well as for simplicity, we consider here solutions with constant topological Euler density.
\par\noindent
The topological Euler density is defined as
\begin{equation}
\mathcal{G}=\frac{1}{32\pi^2}\left(R^{\mu\nu\rho\sigma}R_{\mu\nu\rho\sigma}-4 R^{\mu\nu}R_{\mu\nu}+R^{2} \right).
\end{equation}
%
%
%
For a spherically symmetric solution of the Einstein equations given by Eq. (\ref{metric}) we get
\begin{equation}
\mathcal{G}=\frac{4}{r^2}\left(f'(r)^{2}+\left(f(r)-1\right) f''(r)\right)
\end{equation}
with the factor $32  \pi^2$  to have been included into $\mathcal{G}$ for convenience.
%
%
\subsection{$\mathcal{G}=k\ne 0$ solutions}
%
%
%
%
\par\noindent
The solution of $\mathcal{G}=\nobreak k$ with $k$ to be an arbitrary constant is given by
\begin{equation}
\label{new}
f(r)=1\pm\sqrt{1-2 A+B r +\frac{k r^{4}}{24}}
\end{equation}
where $A$ and $B$ are arbitrary constants. Let us focus on the negative sign case which will give the BH solutions.
\par\noindent
Employing Eq. (\ref{electric}), we get that Eq. (\ref{new}) solves the Einstein-NLED system provided 
the electric field is given by
\begin{widetext}
\begin{equation}
\label{newelectric}
E(r)=\sqrt{6}\; \frac{4 (1-2 A)^2+48 (1-2 A) B r+27 B^2 r^2+(-1+2 A) k r^4}
{q \left(24-48 A+24 B r+k r^4\right)^{3/2}}.
\end{equation}
\end{widetext}
\par\noindent
At this point, a number of comments are in order. 
First,  the electric field asymptotically will be
\begin{equation}
\label{coul}
\lim_{r\rightarrow \infty}E(r) \rightarrow \sqrt{\frac{6}{ k}} \frac{2 A-1}{q r^{2}}
\end{equation}
which corresponds to a Coulomb-like behavior with a  charge $q^{2} = \sqrt{\frac{6}{ k}} (2 A-1)$.
Second, if the constants are   selected to be $A=1/2$ and $B=0$, then Eq. (\ref{new}) corresponds to de Sitter 
or anti de Sitter space and its electric field vanishes, as shown by Eq. (\ref{newelectric}).
Third, it is evident from Eq. (\ref{coul}) that the asymptotic electric field does not depend on $B$. 
Fourth, the metric element as given in Eq. (\ref{new}) asymptotically reads
\begin{equation}
\label{lim}
\lim_{r\rightarrow\infty} f(r)\rightarrow 1-\sqrt{\frac{6}{k}}\frac{B}{r}-\sqrt{\frac{6}{k}}\frac{2A-1}{r^2}-\sqrt{\frac{k}{6}}\frac{r^2}{2}
\end{equation}
which corresponds to a Reissner-Nordstr\"om-de Sitter solution provided that 
\begin{eqnarray}
\label{constants}
A &=&\frac{1}{2}+ \frac{ q^{2} \Lambda }{3} \nonumber \\
B&=& \frac{4M\Lambda}{3} \nonumber \\
k&=&\frac{8\Lambda^{2}}{3}.
\end{eqnarray}
%

%
%
\par\noindent
Therefore, utilizing Eqs. (\ref{new}) and  (\ref{constants}), a BH solution of the coupled Einstein-NLED system 
for a certain electromagnetic Hamiltonian which will be derived in the Appendix, will be of the form
%
\begin{eqnarray}
\label{final}
ds^{2}\!\!\!&=&\!\!\!-\left(1-\sqrt{\frac{4 M \Lambda r}{3}+\frac{\Lambda^{2}}{9}r^{4}-\frac{2 q^{2}\Lambda}{3}}\right)dt^{2}\nonumber\\
\!\!\!&&\!\!\!+\frac{dr^{2}}{\left(1-\sqrt{\frac{4 M \Lambda r}{3}+\frac{\Lambda^{2}}{9}r^{4}-\frac{2 q^{2}\Lambda}{3}}\right)} +r^2 d \Omega^{2} ~.
\end{eqnarray}
%
%
%
\par\noindent
Let us now study  the massless ($M=0$) and massive ($M\ne 0$) cases, separately.
%
%
%
\subsubsection{Massless case}
%
%
%
\par\noindent
In this case, the metric element as given in Eq. (\ref{final}) now reads
%
%
%
%
\begin{eqnarray}
\label{final1}
ds^{2}\!\!\!&=&\!\!\!-\left(1-\sqrt{\frac{\Lambda^{2}}{9}r^{4}-\frac{2 q^{2}\Lambda}{3}}\right)dt^{2}\nonumber\\
\!\!\!&&\!\!\!+\frac{dr^{2}}{\left(1-\sqrt{\frac{\Lambda^{2}}{9}r^{4}-\frac{2 q^{2}\Lambda}{3}}\right)} +r^2 d \Omega^2 ~.
\end{eqnarray}
%
%
%
%
Depending on the values of the parameters, $q$ and $\Lambda$, of this metric, one gets:
\begin{itemize}
\item $q=0$ and  $\Lambda\ne0$\\
This case corresponds to the de Sitter solution which has a cosmological horizon at $r_{c}=\nobreak (3/\Lambda)^{1/2}$.
\item $q\ne0$ and $\Lambda\ne0$\\
This case has an event horizon located at $r_{h}=\left[\frac{9}{\Lambda^2}+\frac{6 q^2}{\Lambda}\right]^{1/4}$
with $\Lambda < - 3 / 2 q^{2}$. In addition, using Eqs. (\ref{newelectric}) and (\ref{constants}), the electric field reads
\begin{equation}
\label{elec}
E(r)=q\frac{r^4 +6 q^2/\Lambda}{\left(r^4-6q^2/\Lambda\right)^{3/2}}
\end{equation}
which asymptotically becomes, as expected, 
\begin{equation}
\lim_{r\rightarrow \infty}E(r) \rightarrow \frac{q}{r^2}~.
\end{equation}
\end{itemize}
\par\noindent
At this point, a number of comments are in order. First,  the metric element, i.e., $f(r)$, of Eq. (\ref{final1}) 
can be written in the form
\begin{eqnarray}
\label{deviation1}
f(r)&=&1-\frac{\Lambda}{3}r^2\sqrt{1-\frac{\alpha}{r^4}}\nonumber \\
&=&1-\frac{\Lambda}{3}r^2+\frac{q^2}{r^2}+\mathcal{O}\left(\alpha^{2}\right)
\end{eqnarray}
\par\noindent
with $\alpha\equiv 6q^2/\Lambda$ to be a parameter that measures the deviation of the geometry described by 
the metric element given in Eq. (\ref{deviation1}) from de Sitter space.
%
%
%
%
This deviation is depicted in Figure 1.
%
%
%
\begin{figure}[h!]
\includegraphics[scale=0.55]{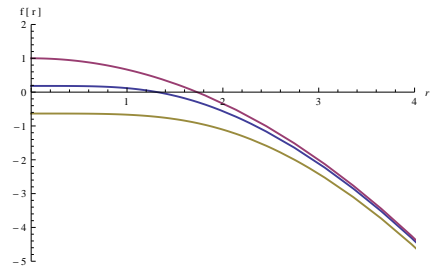}
\caption{The red curve is  the de Sitter space with $\Lambda=-1$, the blue curve is a spacetime with the metric function 
$f(r)$ to have $q=1$ and $\Lambda=-1$, and the olive green curve depicts a spacetime 
with metric function to have $q=2$ and $\Lambda=-1$.}
\label{fig}
\end{figure}
\par\noindent
Second, it is evident from Eq. (\ref{elec}) that there is an  intrinsic singularity located at $r_{s}=\alpha^{1/4}$ 
with $\Lambda>0$. This singularity can also be detected if one computes the curvature invariants 
which, in this case, are given by
\begin{widetext}
\begin{eqnarray}
\label{inv2}
R^{\mu\nu\rho\sigma}R_{\mu\nu\rho\sigma}&=&\frac{8 \Lambda  \left(216 q^8-144 q^6 r^4 \Lambda +114 q^4 r^8 \Lambda ^2-18 q^2 r^{12} \Lambda ^3+r^{16} \Lambda ^4\right)}
{3 r^4 \left(\Lambda r^4  - 6 q^2  \right)^3}
  \nonumber   \\
R^{\mu\nu}R_{\mu\nu}&=&4 \Lambda  \left(\Lambda +\frac{2 q^4 \left(36 q^4+60 q^2 r^4 \Lambda -7 r^8 \Lambda ^2\right)}
{r^4 \left(\Lambda    r^4 - 6 q^2\right)^3}\right)
\nonumber \\
g^{\mu\nu}R_{\mu\nu}&=&\frac{4 \sqrt{\Lambda  \left(-6 q^2+r^4 \Lambda \right)} \left(6 q^4-9 q^2 r^4 \Lambda +r^8 \Lambda ^2\right)}
{r^{2} \left( \Lambda   r^{4}   -  6 q^2 \right)^{2}}.
\end{eqnarray}
\end{widetext}
\par\noindent
It should be stressed that, for $\Lambda>0$, the singularity $r_{s}$ will satisfy $r_{h}>r_{s}$ and, therefore, 
it will be an intrinsic singularity which will never become a naked one. 
Furthermore, this singularity can be avoided by choosing $\Lambda<0$. However,  it is easily seen by inspecting 
the curvature invariants in Eq. (\ref{inv2}) that another singularity is present. 
The singularity at $r=0$ becomes relevant in this case and, therefore, it is not possible a non-singular solution 
to be achieved although the electric field as given by Eq. (\ref{elec}) is regular everywhere. 
It is also noteworthy, that the singularity lying at the origin, i.e., at $r=0$, becomes a naked one when 
$\Lambda\ge -3 / 2 q^{2}$.
%
%
%
%
%
%
\subsubsection{Massive case}
%
%
%
\par\noindent
This is the general case so the metric element is the one  given by Eq. (\ref{final}) 
%
%
%
%
\begin{eqnarray}
\label{final2}
ds^{2}\!\!\!&=&\!\!\!-\left(1-\sqrt{\frac{4 M \Lambda r}{3}+\frac{\Lambda^{2}}{9}r^{4}-\frac{2 q^{2}\Lambda}{3}}\right)dt^{2}\nonumber\\
\!\!\!&&\!\!\!+\frac{dr^{2}}{\left(1-\sqrt{\frac{4 M \Lambda r}{3}+\frac{\Lambda^{2}}{9}r^{4}-\frac{2 q^{2}\Lambda}{3}}\right)} +r^2 d \Omega^{2}~.
\end{eqnarray}
%
%
%
%
\par\noindent
Several comments for this general case have been given at the beginning of this section. In the case that 
$q\ne 0$ and $\Lambda=0$, the geometry is that of Minskowski spacetime. Therefore, the case which will be studied now
is the one with $q=0$ and $\Lambda\ne0$. The corresponding curvature invariants are written as
\begin{widetext}
\begin{eqnarray}
\label{inv3}
R^{\mu\nu\rho\sigma}R_{\mu\nu\rho\sigma}&=& \frac{8}{3} \Lambda  \left(\Lambda +\frac{4374 M^4}{\left(12 M r+r^4 \Lambda \right)^3}\right)  \nonumber   \\
R^{\mu\nu}R_{\mu\nu}&=& \frac{4 \Lambda  \left(2754 M^4+1620 M^3 r^3 \Lambda +414 M^2 r^6 \Lambda ^2+36 M r^9 \Lambda ^3+r^{12} \Lambda ^4\right)}{\left(12 M r+r^4 \Lambda \right)^3} \nonumber \\
g^{\mu\nu}R_{\mu\nu}&=&\frac{4 \left(3 M+r^3 \Lambda \right) \sqrt{r \Lambda  \left(12 M+r^3 \Lambda \right)} \left(15 M+r^3 \Lambda \right)}{\left(12 M r+r^4 
\Lambda \right)^2}~.
\end{eqnarray}
\end{widetext}
\par\noindent
Here, a number of comments are in order.
First, it is evident from Eq.  (\ref{inv3}) that provided $\Lambda>0$ the solution is regular everywhere except at $r=0$.
Second, as no electric charge, i.e., $q=0$, is present in this case, the interpretation
in terms of certain NLED is not of any interest. However, the corresponding source can be taken to be a Type II perfect
fluid whose density and pressure are given, respectively, as
\begin{eqnarray}
\label{rho+pres}
\rho &=& \frac{\Lambda  \left(6 M+r^3 \Lambda \right)}{8 \pi  r \sqrt{r \Lambda  \left(12 M+r^3 \Lambda \right)}} \nonumber  \\
p &=&-\frac{\Lambda ^2 \left(18 M^2+18 M r^3 \Lambda +r^6 \Lambda ^2\right)}{8 \pi  \left[r \Lambda  \left(12 M+r^3 \Lambda \right)\right]^{3/2}}~.
\end{eqnarray}
Third, if $M\ll \Lambda$, then  the corresponding spacetime will be a de Sitter  one  
since the density and the pressure become, respectively, 
\begin{eqnarray}
\rho &=& \frac{\Lambda}{8 \pi } \left( 1 + \mathcal{O}\left( (M/ \Lambda)^{2}\right) \right )\nonumber  \\
p &=& - \left[1 - \frac{270 M^2}{r^6 \Lambda ^2}+  \mathcal{O}\left( (M/ \Lambda)^{4} \right) \right]\rho ~.
\end{eqnarray}
Moreover,  the pressure can also be written as
\begin{equation}
 p=- \left[1 - \frac{270 M^2}{\Lambda ^2 r^6}+\mathcal{O}\left(\frac{1}{r^9}\right)\right]\rho
\end{equation}
\par\noindent
and, thus, the geometry becomes de Sitter also at spatial infinity.
\par\noindent
Fourth, this solution satisfies both the weak and dominant energy conditions while 
the strong energy condition is violated due to the near-de Sitter behavior, as expected.\\
%
%
%
%
%
%
%
%
\subsection{$\mathcal{G}=k=0$ solutions}
%
%
%
%
%
%
%
\par\noindent
The solution of $\mathcal{G}=\nobreak k = 0$ is given by
\begin{equation}
\label{newB}
f(r)=1\pm\sqrt{1-2 A+B r }~.
\end{equation}
We focus, as before,  on the negative sign case which will give the BH solutions.
In this case, the mass function reads
\begin{equation}
\label{mass}
\mathcal{M}(r)=\frac{r}{2}\sqrt{1-2 A + B r}
\end{equation}
and the curvature invariants become
\begin{widetext}
\begin{eqnarray}
\label{inv4}
&&R^{\mu\nu\rho\sigma}R_{\mu\nu\rho\sigma}=\frac{\left(8+32 A^2-32 A (1+r B)+r B (16+9 r B)\right)^2}{16 r^4 (1-2 A+r B)^3} \nonumber   \\
&&R^{\mu\nu}R_{\mu\nu}= \frac{64 (1-2 A)^4-320 r (-1+2 A)^3 B+608 r^2 (1-2 A)^2 B^2+504 r^3 (1-2 A) B^3+153 r^4 B^4}{32 r^4 (1-2 A+r B)^3}\nonumber \\
&&R=\frac{8+32 A^2-16 A (2+3 r B)+3 r B (8+5 r B)}{4 r^2 (1-2 A+r B)^{3/2}}~.
\end{eqnarray}
\end{widetext}
\par\noindent
Depending on the values of the constants, $A$ and $B$, the following cases are considered:
\begin{itemize}
\item $A=1/2$, $B=0$\\ 
In this case, the geometry corresponds to Minkowski spacetime and 
the three energy conditions are trivially satisfied.
\item $A<1/2$ and $B=0$\\
In this case, there is a naked singularity located at $r=0$. In addition, the density and 
the pressure are given, respectively, as 
\begin{eqnarray}
\rho&=&\frac{\sqrt{1-2 A}}{8 \pi  r^2} \nonumber \\
p&=&0~.
\end{eqnarray}
Therefore, the weak, strong, and dominant energy conditions are satisfied provided $A<1/2$. 
Furthermore, this case corresponds to that of the gravitational field
of a global monopole \cite{Barriola1989,Wang1999} with a deficit angle given by $\Delta=\sqrt{1-2A}$.
\item $A=1/2$, $B> 0$\\ 
In this case, there is a horizon at $r=B^{-1}$ surrounding a singularity at $r=0$. In addition, 
the density and the pressure are given, respectively, as
\begin{eqnarray}
\rho&=&\frac{3 \sqrt{B r}}{16 \pi  r^2} \nonumber \\
p&=&-\frac{3 \sqrt{B r}}{64 \pi  r^2}~.
\end{eqnarray}
\par\noindent
Therefore, the weak and dominant,  but not the strong,  energy conditions are satisfied. It is noteworthy that 
this situation corresponds to a particular case of quintessence dark matter (see, for example, Ref. \cite{Ellis2012}) 
whose equation of state is of the form $p=-\frac{1}{4}\rho$.
\item $A\ne0$, $B\ne0$\\ 
In this general case, the density and pressure are given, respectively, as
\begin{eqnarray}
\label{general}
\rho&=&\frac{2-4 A+3 B r}{16 \pi  r^2 \sqrt{1-2 A+B r}}\nonumber \\
p&=&-\frac{B (4-8 A+3 B r)}{64 \pi  r (1-2 A+B r)^{3/2}}.
\end{eqnarray}
%
%
%
%
\par\noindent
It is easily seen that the most interesting case is when $0<2A<1$ and $B>0$. In this situation,  
there is a horizon at $r=2A/B$ surrounding the singularity at $r=0$. Both the weak and dominant energy conditions 
are satisfied when $r>0$. On the contrary, the strong energy condition is never fulfilled.
\end{itemize}
%
%
%
%
%
%
%
%
\section{Discussion and conclusions}
%
%
%
\par\noindent
In this work we have studied a class of four dimensional spherically symmetric and static geometries with constant 
topological Euler density, i.e., $\mathcal{G}$, showing that they can be interpreted as Reissner-Nordstr\"om-de Sitter-like 
spacetimes when non-linear electrodynamics is utilized, or as generalized Vaidya solutions, depending on the value 
of $\mathcal{G}$.
\par\noindent
In the first case in which the topological Euler density is a non-zero constant, i.e., $\mathcal{G}\ne0$, 
we managed to show for the massless case, i.e., $M=0$, that the non-linear electric field
can be regularized everywhere by taking $\Lambda<0$, although
the geometry remains singular at $r=0$. Furthermore, it was shown that when $\Lambda \ge -3/2 q^{2} $, 
the singularity lying at the origin, i.e., $r=0$, becomes naked. 
For the massive case, i.e., $M\ne 0$ when the charge is switched off, then provided that $\Lambda > 0$, the obtained geometry is
regular everywhere except the origin,  i.e., $r=0$, where a singularity lies.  In this situation, the source term is interpreted in terms of generalized Vaidya geometries. 
Contrary to the previous case, now both the density and pressure of the fluid become singular at $r=0$. 
\par\noindent
In the second case in which the topological Euler density is equal to zero, i.e., $\mathcal{G}=0$, the geometries obtained are 
that of Minkowski spacetime, of a global monopole, of a quintessence dark matter model, and of a BH with an horizon 
surrounding the singularity at $r=0$. 
\par\noindent
Finally, it should be stressed that, as stated in \cite{Garcia2012}, the only kinds of matter consistent with
spherically symmetric and static gravitational fields in General Relativity are the cosmological constant vacuum, 
the anisotropic fluids, the perfect fluids,  and the linear/non-linear Abelian/non-Abelian electromagnetic/Yang-Mills fields. 
Interestingly, as it was pointed out in Ref. \cite{Garcia2012}, this latter case also allows the interpretation of 
an anisotropic fluid in certain cases, one of which has been studied in the present work.
%
%
%
%
\section{Acknowledgments}
\par\noindent
P. B. acknowledges support from the Faculty of Science and Vicerrector\'{\i}a de Investigaciones of
Universidad de los Andes, Bogot\'a, Colombia.\\
%
%
%
\section{Appendix}
%
%
%
\par\noindent
Using the dual formalism briefly described in section II, the underlying theory is shown to be given by
\begin{equation}
\label{NLED}
\mathcal{H}(P) =\left( -\frac{3}{2}\tilde{a}q +\frac{\tilde{b}P}{q} \right)\frac{1}{\sqrt{\tilde{a} q^{2}- 2 \tilde{b}P}}
\end{equation}
with $\tilde{a}=\frac{\Lambda^{2}}{q}$ and $\tilde{b}=\frac{2}{3}q^{2}\Lambda$.
Let us note that for small fields ($P\ll \Lambda$)
\begin{equation}
\label{lagrangian}
\mathcal{H}(P) =- \frac{\Lambda}{2} - P +12\sqrt{\Lambda}\, P^{2}+\mathcal{O}(P^3 )~.
\end{equation}
\par\noindent
%
%
%
\par\noindent
At this point, it is worth of note that the BI Hamiltonian when expanded for small fields compared to 
the maximal field strength, $b$, reads 
\begin{equation}
\mathcal{H}_{BI}=-F+\frac{F^2}{2 b^2}+\mathcal{O}\left(F^{3}\right)~.
\end{equation}
%
%
%
\par\noindent
Therefore, the NLED model, described here, gives place to a BI-like model (up to $F^2$) 
when $ b^{2}= \frac{1}{24\sqrt{\Lambda}}$. 
%
%
%

\end{document}